\pdfoutput=1
\documentclass[conference,a4paper]{IEEEtran}
\IEEEoverridecommandlockouts
\usepackage{cite}
\usepackage{amsmath,amssymb,amsfonts}
\usepackage{algorithmic}
\usepackage{graphicx}
\usepackage{textcomp}
\usepackage{xcolor}
\usepackage{listofitems}
\usepackage{todonotes}
\def\BibTeX{{\rm B\kern-.05em{\sc i\kern-.025em b}\kern-.08em
    T\kern-.1667em\lower.7ex\hbox{E}\kern-.125emX}}

\usepackage[acronyms,nonumberlist,nopostdot,nomain,nogroupskip,acronymlists={hidden}]{glossaries}
\newglossary[algh]{hidden}{acrh}{acnh}{Hidden Acronyms}

\usepackage{booktabs}
\usepackage{tabularx}

\usepackage{tikz}
\usepackage{pgfplots}
\usepackage{subcaption}
\pgfplotsset{compat=newest}
\pgfplotsset{plot coordinates/math parser=false}
\newlength\fheight
\newlength\fwidth
\usetikzlibrary{plotmarks,patterns,decorations.pathreplacing,backgrounds,calc,arrows,arrows.meta,spy,matrix,scopes}
\usepgfplotslibrary{patchplots,groupplots,units}
\usetikzlibrary{pgfplots.statistics, pgfplots.colorbrewer} 
\usepackage{pgfplotstable}
\usepackage{tikzscale}
\usepackage[draft]{hyperref}
\usepackage{siunitx}

\newif\ifexttikz
\exttikzfalse

\ifexttikz
	\usetikzlibrary{external}
\fi

\usepackage{multirow}

\usepackage[font=footnotesize]{subcaption}
\usepackage[font=footnotesize]{caption}

\usepackage{mathtools}
\usepackage[numbers,sort&compress]{natbib}
\usepackage{soul}
\usepackage{cleveref}
\crefname{section}{Sect.}{Sects.}
\crefname{figure}{Fig.}{Figs.}
\crefname{table}{Tab.}{Tabs.}
\crefname{equation}{Eq.}{Eqs.}

\newacronym{3gpp}{3GPP}{3rd Generation Partnership Project}
\newacronym{4g}{4G}{4th generation}
\newacronym{5g}{5G}{5th generation}
\newacronym{6g}{6G}{6th generation}
\newacronym{5gc}{5GC}{5G Core}
\newacronym{adc}{ADC}{Analog to Digital Converter}
\newacronym{aerpaw}{AERPAW}{Aerial Experimentation and Research Platform for Advanced Wireless}
\newacronym{ai}{AI}{Artificial Intelligence}
\newacronym{aimd}{AIMD}{Additive Increase Multiplicative Decrease}
\newacronym{am}{AM}{Acknowledged Mode}
\newacronym{amc}{AMC}{Adaptive Modulation and Coding}
\newacronym{amf}{AMF}{Access and Mobility Management Function}
\newacronym{aops}{AOPS}{Adaptive Order Prediction Scheduling}
\newacronym{api}{API}{Application Programming Interface}
\newacronym{apn}{APN}{Access Point Name}
\newacronym{ap}{AP}{Application Protocol}
\newacronym{aqm}{AQM}{Active Queue Management}
\newacronym{ausf}{AUSF}{Authentication Server Function}
\newacronym{avc}{AVC}{Advanced Video Coding}
\newacronym{awgn}{AGWN}{Additive White Gaussian Noise}
\newacronym{balia}{BALIA}{Balanced Link Adaptation Algorithm}
\newacronym{bbu}{BBU}{Base Band Unit}
\newacronym{bdp}{BDP}{Bandwidth-Delay Product}
\newacronym{ber}{BER}{Bit Error Rate}
\newacronym{bf}{BF}{Beamforming}
\newacronym{bler}{BLER}{Block Error Rate}
\newacronym{brr}{BRR}{Bayesian Ridge Regressor}
\newacronym{bs}{BS}{Base Station}
\newacronym{bsr}{BSR}{Buffer Status Report}
\newacronym{bss}{BSS}{Business Support System}
\newacronym{ca}{CA}{Carrier Aggregation}
\newacronym{caas}{CaaS}{Connectivity-as-a-Service}
\newacronym{cb}{CB}{Code Block}
\newacronym{cc}{CC}{Congestion Control}
\newacronym{ccid}{CCID}{Congestion Control ID}
\newacronym{cco}{CC}{Carrier Component}
\newacronym{cd}{CD}{Continuous Delivery}
\newacronym{cdd}{CDD}{Cyclic Delay Diversity}
\newacronym{cdf}{CDF}{Cumulative Distribution Function}
\newacronym{cdn}{CDN}{Content Distribution Network}
\newacronym{cli}{CLI}{Command-line Interface}
\newacronym{cn}{CN}{Core Network}
\newacronym{codel}{CoDel}{Controlled Delay Management}
\newacronym{comac}{COMAC}{Converged Multi-Access and Core}
\newacronym{cord}{CORD}{Central Office Re-architected as a Datacenter}
\newacronym{cornet}{CORNET}{COgnitive Radio NETwork}
\newacronym{cosmos}{COSMOS}{Cloud Enhanced Open Software Defined Mobile Wireless Testbed for City-Scale Deployment}
\newacronym{cots}{COTS}{Commercial Off-the-Shelf}
\newacronym{cp}{CP}{Control Plane}
\newacronym{cyp}{CP}{Cyclic Prefix}
\newacronym{up}{UP}{User Plane}
\newacronym{cpu}{CPU}{Central Processing Unit}
\newacronym{cqi}{CQI}{Channel Quality Information}
\newacronym{cr}{CR}{Cognitive Radio}
\newacronym{cran}{CRAN}{Cloud \gls{ran}}
\newacronym{crs}{CRS}{Cell Reference Signal}
\newacronym{csi}{CSI}{Channel State Information}
\newacronym{csirs}{CSI-RS}{Channel State Information - Reference Signal}
\newacronym{cu}{CU}{Central Unit}
\newacronym{d2tcp}{D$^2$TCP}{Deadline-aware Data center TCP}
\newacronym{d3}{D$^3$}{Deadline-Driven Delivery}
\newacronym{dac}{DAC}{Digital to Analog Converter}
\newacronym{dag}{DAG}{Directed Acyclic Graph}
\newacronym{das}{DAS}{Distributed Antenna System}
\newacronym{dash}{DASH}{Dynamic Adaptive Streaming over HTTP}
\newacronym{dc}{DC}{Dual Connectivity}
\newacronym{dccp}{DCCP}{Datagram Congestion Control Protocol}
\newacronym{dce}{DCE}{Direct Code Execution}
\newacronym{dci}{DCI}{Downlink Control Information}
\newacronym{dctcp}{DCTCP}{Data Center TCP}
\newacronym{dl}{DL}{Downlink}
\newacronym{dmr}{DMR}{Deadline Miss Ratio}
\newacronym{dmrs}{DMRS}{DeModulation Reference Signal}
\newacronym{drlcc}{DRL-CC}{Deep Reinforcement Learning Congestion Control}
\newacronym{drs}{DRS}{Discovery Reference Signal}
\newacronym{du}{DU}{Distributed Unit}
\newacronym{e2e}{E2E}{end-to-end}
\newacronym{earfcn}{EARFCN}{E-UTRA Absolute Radio Frequency Channel Number}
\newacronym{ecaas}{ECaaS}{Edge-Cloud-as-a-Service}
\newacronym{ecn}{ECN}{Explicit Congestion Notification}
\newacronym{edf}{EDF}{Earliest Deadline First}
\newacronym{embb}{eMBB}{Enhanced Mobile Broadband}
\newacronym{empower}{EMPOWER}{EMpowering transatlantic PlatfOrms for advanced WirEless Research}
\newacronym{enb}{eNB}{evolved Node Base}
\newacronym{endc}{EN-DC}{E-UTRAN-\gls{nr} \gls{dc}}
\newacronym{epc}{EPC}{Evolved Packet Core}
\newacronym{eps}{EPS}{Evolved Packet System}
\newacronym{es}{ES}{Edge Server}
\newacronym{etsi}{ETSI}{European Telecommunications Standards Institute}
\newacronym[firstplural=Estimated Times of Arrival (ETAs)]{eta}{ETA}{Estimated Time of Arrival}
\newacronym{eutran}{E-UTRAN}{Evolved Universal Terrestrial Access Network}
\newacronym{faas}{FaaS}{Function-as-a-Service}
\newacronym{fapi}{FAPI}{Functional Application Platform Interface}
\newacronym{fdd}{FDD}{Frequency Division Duplexing}
\newacronym{fdm}{FDM}{Frequency Division Multiplexing}
\newacronym{fdma}{FDMA}{Frequency Division Multiple Access}
\newacronym{fed4fire}{FED4FIRE+}{Federation 4 Future Internet Research and Experimentation Plus}
\newacronym{fir}{FIR}{Finite Impulse Response}
\newacronym{fit}{FIT}{Future \acrlong{iot}}
\newacronym{fpga}{FPGA}{Field Programmable Gate Array}
\newacronym{fr2}{FR2}{Frequency Range 2}
\newacronym{fs}{FS}{Fast Switching}
\newacronym{fscc}{FSCC}{Flow Sharing Congestion Control}
\newacronym{ftp}{FTP}{File Transfer Protocol}
\newacronym{fw}{FW}{Flow Window}
\newacronym{ge}{GE}{Gaussian Elimination}
\newacronym{gnb}{gNB}{Next Generation Node Base}
\newacronym{gop}{GOP}{Group of Pictures}
\newacronym{gpr}{GPR}{Gaussian Process Regressor}
\newacronym{gpu}{GPU}{Graphics Processing Unit}
\newacronym{gtp}{GTP}{GPRS Tunneling Protocol}
\newacronym{gtpc}{GTP-C}{GPRS Tunnelling Protocol Control Plane}
\newacronym{gtpu}{GTP-U}{GPRS Tunnelling Protocol User Plane}
\newacronym{gtpv2c}{GTPv2-C}{\gls{gtp} v2 - Control}
\newacronym{gw}{GW}{Gateway}
\newacronym{harq}{HARQ}{Hybrid Automatic Repeat reQuest}
\newacronym{hetnet}{HetNet}{Heterogeneous Network}
\newacronym{hh}{HH}{Hard Handover}
\newacronym{hol}{HOL}{Head-of-Line}
\newacronym{hqf}{HQF}{Highest-quality-first}
\newacronym{hss}{HSS}{Home Subscription Server}
\newacronym{http}{HTTP}{HyperText Transfer Protocol}
\newacronym{ia}{IA}{Initial Access}
\newacronym{iab}{IAB}{Integrated Access and Backhaul}
\newacronym{ic}{IC}{Incident Command}
\newacronym{ietf}{IETF}{Internet Engineering Task Force}
\newacronym{imsi}{IMSI}{International Mobile Subscriber Identity}
\newacronym{imt}{IMT}{International Mobile Telecommunication}
\newacronym{iot}{IoT}{Internet of Things}
\newacronym{ip}{IP}{Internet Protocol}
\newacronym{itu}{ITU}{International Telecommunication Union}
\newacronym{kpi}{KPI}{Key Performance Indicator}
\newacronym{kpm}{KPM}{Key Performance Measurement}
\newacronym{kvm}{KVM}{Kernel-based Virtual Machine}
\newacronym{los}{LoS}{Line of Sight}
\newacronym{lsm}{LSM}{Link-to-System Mapping}
\newacronym{lstm}{LSTM}{Long Short Term Memory}
\newacronym{lte}{LTE}{Long Term Evolution}
\newacronym{lxc}{LXC}{Linux Container}
\newacronym{m2m}{M2M}{Machine to Machine}
\newacronym{mac}{MAC}{Medium Access Control}
\newacronym{manet}{MANET}{Mobile Ad Hoc Network}
\newacronym{mano}{MANO}{Management and Orchestration}
\newacronym{mc}{MC}{Multi-Connectivity}
\newacronym{mcc}{MCC}{Mobile Cloud Computing}
\newacronym{mchem}{MCHEM}{Massive Channel Emulator}
\newacronym{mcs}{MCS}{Modulation and Coding Scheme}
\newacronym{mec2}{MEC}{Multi-access Edge Computing}
\newacronym{mec}{MEC}{Mobile Edge Computing}
\newacronym{mfc}{MFC}{Mobile Fog Computing}
\newacronym{mgen}{MGEN}{Multi-Generator}
\newacronym{mi}{MI}{Mutual Information}
\newacronym{mib}{MIB}{Master Information Block}
\newacronym{miesm}{MIESM}{Mutual Information Based Effective SINR}
\newacronym{mimo}{MIMO}{Multiple Input, Multiple Output}
\newacronym{ml}{ML}{Machine Learning}
\newacronym{mlr}{MLR}{Maximum-local-rate}
\newacronym[plural=\gls{mme}s,firstplural=Mobility Management Entities (MMEs)]{mme}{MME}{Mobility Management Entity}
\newacronym{mmtc}{mMTC}{Massive Machine-Type Communications}
\newacronym{mmwave}{mmWave}{millimeter wave}
\newacronym{mpdccp}{MP-DCCP}{Multipath Datagram Congestion Control Protocol}
\newacronym{mptcp}{MPTCP}{Multipath TCP}
\newacronym{mr}{MR}{Maximum Rate}
\newacronym{mrdc}{MR-DC}{Multi \gls{rat} \gls{dc}}
\newacronym{mse}{MSE}{Mean Square Error}
\newacronym{mss}{MSS}{Maximum Segment Size}
\newacronym{mt}{MT}{Mobile Termination}
\newacronym{mtd}{MTD}{Machine-Type Device}
\newacronym{mtu}{MTU}{Maximum Transmission Unit}
\newacronym{mumimo}{MU-MIMO}{Multi-user \gls{mimo}}
\newacronym{mvno}{MVNO}{Mobile Virtual Network Operator}
\newacronym{nalu}{NALU}{Network Abstraction Layer Unit}
\newacronym{nas}{NAS}{Network Attached Storage}
\newacronym{nat}{NAT}{Network Address Translation}
\newacronym{nbiot}{NB-IoT}{Narrow Band IoT}
\newacronym{nfv}{NFV}{Network Function Virtualization}
\newacronym{nfvi}{NFVI}{Network Function Virtualization Infrastructure}
\newacronym{ni}{NI}{Network Interfaces}
\newacronym{nic}{NIC}{Network Interface Card}
\newacronym{now}{NOW}{Non Overlapping Window}
\newacronym{nsm}{NSM}{Network Service Mesh}
\newacronym{nr}{NR}{New Radio}
\newacronym{nrf}{NRF}{Network Repository Function}
\newacronym{nsa}{NSA}{Non Stand Alone}
\newacronym{nse}{NSE}{Network Slicing Engine}
\newacronym{nssf}{NSSF}{Network Slice Selection Function}
\newacronym{o2i}{O2I}{Outdoor to Indoor}
\newacronym{oai}{OAI}{OpenAirInterface}
\newacronym{oaicn}{OAI-CN}{\gls{oai} \acrlong{cn}}
\newacronym{oairan}{OAI-RAN}{\acrlong{oai} \acrlong{ran}}
\newacronym{oam}{OAM}{Operations, Administration and Maintenance}
\newacronym{ofdm}{OFDM}{Orthogonal Frequency Division Multiplexing}
\newacronym{olia}{OLIA}{Opportunistic Linked Increase Algorithm}
\newacronym{omec}{OMEC}{Open Mobile Evolved Core}
\newacronym{onap}{ONAP}{Open Network Automation Platform}
\newacronym{onf}{ONF}{Open Networking Foundation}
\newacronym{onos}{ONOS}{Open Networking Operating System}
\newacronym{oom}{OOM}{\gls{onap} Operations Manager}
\newacronym{opnfv}{OPNFV}{Open Platform for \gls{nfv}}
\newacronym{oran}{O-RAN}{Open \gls{ran}}
\newacronym{orbit}{ORBIT}{Open-Access Research Testbed for Next-Generation Wireless Networks}
\newacronym{os}{OS}{Operating System}
\newacronym{oss}{OSS}{Operations Support System}
\newacronym{pa}{PA}{Position-aware}
\newacronym{pase}{PASE}{Prioritization, Arbitration, and Self-adjusting Endpoints}
\newacronym{pawr}{PAWR}{Platforms for Advanced Wireless Research}
\newacronym{pbch}{PBCH}{Physical Broadcast Channel}
\newacronym{pcef}{PCEF}{Policy and Charging Enforcement Function}
\newacronym{pcfich}{PCFICH}{Physical Control Format Indicator Channel}
\newacronym{pcrf}{PCRF}{Policy and Charging Rules Function}
\newacronym{pdcch}{PDCCH}{Physical Downlink Control Channel}
\newacronym{pdcp}{PDCP}{Packet Data Convergence Protocol}
\newacronym{pdsch}{PDSCH}{Physical Downlink Shared Channel}
\newacronym{pdu}{PDU}{Packet Data Unit}
\newacronym{pf}{PF}{Proportional Fair}
\newacronym{pgw}{PGW}{Packet Gateway}
\newacronym{phich}{PHICH}{Physical Hybrid ARQ Indicator Channel}
\newacronym{phy}{PHY}{Physical}
\newacronym{pmch}{PMCH}{Physical Multicast Channel}
\newacronym{pmi}{PMI}{Precoding Matrix Indicators}
\newacronym{powder}{POWDER}{Platform for Open Wireless Data-driven Experimental Research}
\newacronym{ppo}{PPO}{Proximal Policy Optimization}
\newacronym{ppp}{PPP}{Poisson Point Process}
\newacronym{prach}{PRACH}{Physical Random Access Channel}
\newacronym{prb}{PRB}{Physical Resource Block}
\newacronym{psnr}{PSNR}{Peak Signal to Noise Ratio}
\newacronym{pss}{PSS}{Primary Synchronization Signal}
\newacronym{pucch}{PUCCH}{Physical Uplink Control Channel}
\newacronym{pusch}{PUSCH}{Physical Uplink Shared Channel}
\newacronym{qam}{QAM}{Quadrature Amplitude Modulation}
\newacronym{qci}{QCI}{\gls{qos} Class Identifier}
\newacronym{qoe}{QoE}{Quality of Experience}
\newacronym{qos}{QoS}{Quality of Service}
\newacronym{quic}{QUIC}{Quick UDP Internet Connections}
\newacronym{rach}{RACH}{Random Access Channel}
\newacronym{ran}{RAN}{Radio Access Network}
\newacronym[firstplural=Radio Access Technologies (RATs)]{rat}{RAT}{Radio Access Technology}
\newacronym{rbg}{RBG}{Resource Block Group}
\newacronym{rcn}{RCN}{Research Coordination Network}
\newacronym{rc}{RC}{RAN Control}
\newacronym{rec}{REC}{Radio Edge Cloud}
\newacronym{red}{RED}{Random Early Detection}
\newacronym{renew}{RENEW}{Reconfigurable Eco-system for Next-generation End-to-end Wireless}
\newacronym{rf}{RF}{Radio Frequency}
\newacronym{rfc}{RFC}{Request for Comments}
\newacronym{rfr}{RFR}{Random Forest Regressor}
\newacronym{ric}{RIC}{\gls{ran} Intelligent Controller}
\newacronym{rlc}{RLC}{Radio Link Control}
\newacronym{rlf}{RLF}{Radio Link Failure}
\newacronym{rlnc}{RLNC}{Random Linear Network Coding}
\newacronym{rmr}{RMR}{RIC Message Router}
\newacronym{rmse}{RMSE}{Root Mean Squared Error}
\newacronym{rnis}{RNIS}{Radio Network Information Service}
\newacronym{rr}{RR}{Round Robin}
\newacronym{rrc}{RRC}{Radio Resource Control}
\newacronym{rrm}{RRM}{Radio Resource Management}
\newacronym{rru}{RRU}{Remote Radio Unit}
\newacronym{rs}{RS}{Remote Server}
\newacronym{rsrp}{RSRP}{Reference Signal Received Power}
\newacronym{rsrq}{RSRQ}{Reference Signal Received Quality}
\newacronym{rss}{RSS}{Received Signal Strength}
\newacronym{rssi}{RSSI}{Received Signal Strength Indicator}
\newacronym{rtt}{RTT}{Round Trip Time}
\newacronym{ru}{RU}{Radio Unit}
\newacronym{rw}{RW}{Receive Window}
\newacronym{rx}{RX}{Receiver}
\newacronym{s1ap}{S1AP}{S1 Application Protocol}
\newacronym{sa}{SA}{standalone}
\newacronym{sack}{SACK}{Selective Acknowledgment}
\newacronym{sap}{SAP}{Service Access Point}
\newacronym{sc2}{SC2}{Spectrum Collaboration Challenge}
\newacronym{scef}{SCEF}{Service Capability Exposure Function}
\newacronym{sch}{SCH}{Secondary Cell Handover}
\newacronym{scoot}{SCOOT}{Split Cycle Offset Optimization Technique}
\newacronym{sctp}{SCTP}{Stream Control Transmission Protocol}
\newacronym{sdap}{SDAP}{Service Data Adaptation Protocol}
\newacronym{sdk}{SDK}{Software Development Kit}
\newacronym{sdm}{SDM}{Space Division Multiplexing}
\newacronym{sdma}{SDMA}{Spatial Division Multiple Access}
\newacronym{sdn}{SDN}{Software-defined Networking}
\newacronym{sdr}{SDR}{Software-defined Radio}
\newacronym{seba}{SEBA}{SDN-Enabled Broadband Access}
\newacronym{sgsn}{SGSN}{Serving GPRS Support Node}
\newacronym{sgw}{SGW}{Service Gateway}
\newacronym{si}{SI}{Study Item}
\newacronym{sib}{SIB}{Secondary Information Block}
\newacronym{sinr}{SINR}{Signal to Interference plus Noise Ratio}
\newacronym{sip}{SIP}{Session Initiation Protocol}
\newacronym{siso}{SISO}{Single Input, Single Output}
\newacronym{sla}{SLA}{Service Level Agreement}
\newacronym{sm}{SM}{Service Model}
\newacronym{smo}{SMO}{Service Management and Orchestration}
\newacronym{smsgmsc}{SMS-GMSC}{\gls{sms}-Gateway}
\newacronym{snr}{SNR}{Signal-to-Noise-Ratio}
\newacronym{son}{SON}{Self-Organizing Network}
\newacronym{sptcp}{SPTCP}{Single Path TCP}
\newacronym{srb}{SRB}{Service Radio Bearer}
\newacronym{srn}{SRN}{Standard Radio Node}
\newacronym{srs}{SRS}{Sounding Reference Signal}
\newacronym{ss}{SS}{Synchronization Signal}
\newacronym{sss}{SSS}{Secondary Synchronization Signal}
\newacronym{st}{ST}{Spanning Tree}
\newacronym{svc}{SVC}{Scalable Video Coding}
\newacronym{tb}{TB}{Transport Block}
\newacronym{tcp}{TCP}{Transmission Control Protocol}
\newacronym{tdd}{TDD}{Time Division Duplexing}
\newacronym{tdm}{TDM}{Time Division Multiplexing}
\newacronym{tdma}{TDMA}{Time Division Multiple Access}
\newacronym{tfl}{TfL}{Transport for London}
\newacronym{tfrc}{TFRC}{TCP-Friendly Rate Control}
\newacronym{tft}{TFT}{Traffic Flow Template}
\newacronym{tgen}{TGEN}{Traffic Generator}
\newacronym{tip}{TIP}{Telecom Infra Project}
\newacronym{tm}{TM}{Transparent Mode}
\newacronym{to}{TO}{Telco Operator}
\newacronym{tr}{TR}{Technical Report}
\newacronym{trp}{TRP}{Transmitter Receiver Pair}
\newacronym{ts}{TS}{Technical Specification}
\newacronym{tti}{TTI}{Transmission Time Interval}
\newacronym{ttt}{TTT}{Time-to-Trigger}
\newacronym{tx}{TX}{Transmitter}
\newacronym{uas}{UAS}{Unmanned Aerial System}
\newacronym{uav}{UAV}{Unmanned Aerial Vehicle}
\newacronym{udm}{UDM}{Unified Data Management}
\newacronym{udp}{UDP}{User Datagram Protocol}
\newacronym{udr}{UDR}{Unified Data Repository}
\newacronym{ue}{UE}{User Equipment}
\newacronym{uhd}{UHD}{\gls{usrp} Hardware Driver}
\newacronym{ul}{UL}{Uplink}
\newacronym{um}{UM}{Unacknowledged Mode}
\newacronym{uml}{UML}{Unified Modeling Language}
\newacronym{upa}{UPA}{Uniform Planar Array}
\newacronym{upf}{UPF}{User Plane Function}
\newacronym{urllc}{URLLC}{Ultra Reliable and Low Latency Communications}
\newacronym{usa}{U.S.}{United States}
\newacronym{usim}{USIM}{Universal Subscriber Identity Module}
\newacronym{usrp}{USRP}{Universal Software Radio Peripheral}
\newacronym{utc}{UTC}{Urban Traffic Control}
\newacronym{vim}{VIM}{Virtualization Infrastructure Manager}
\newacronym{vm}{VM}{Virtual Machine}
\newacronym{vnf}{VNF}{Virtual Network Function}
\newacronym{volte}{VoLTE}{Voice over \gls{lte}}
\newacronym{voltha}{VOLTHA}{Virtual OLT HArdware Abstraction}
\newacronym{vr}{VR}{Virtual Reality}
\newacronym{vran}{vRAN}{Virtualized \gls{ran}}
\newacronym{vss}{VSS}{Video Streaming Server}
\newacronym{wbf}{WBF}{Wired Bias Function}
\newacronym{wf}{WF}{Waterfilling}
\newacronym{wg}{WG}{Working Group}
\newacronym{wlan}{WLAN}{Wireless Local Area Network}
\newacronym{osm}{OSM}{Open Source \gls{nfv} Management and Orchestration}
\newacronym{pnf}{PNF}{Physical Network Function}
\newacronym{drl}{DRL}{Deep Reinforcement Learning}
\newacronym{mtc}{MTC}{Machine-type Communications}
\newacronym{osc}{OSC}{O-RAN Software Community}
\newacronym{mns}{MnS}{Management Services}
\newacronym{ves}{VES}{\gls{vnf} Event Stream}
\newacronym{ei}{EI}{Enrichment Information}
\newacronym{fh}{FH}{Fronthaul}
\newacronym{fft}{FFT}{Fast Fourier Transform}
\newacronym{laa}{LAA}{Licensed-Assisted Access}
\newacronym{plfs}{PLFS}{Physical Layer Frequency Signals}
\newacronym{ptp}{PTP}{Precision Time Protocol}
\newacronym{lidar}{LiDAR}{Light Detection And Ranging}
\newacronym{dem}{DEM}{Digital Elevation Model}
\newacronym{dtm}{DEM}{Digital Terrain Model}
\newacronym{dsm}{DEM}{Digital Surface Models}
\newacronym{ota}{OTA}{Over-The-Air}
\newacronym{ns}{NS}{Network Slicing}
\newacronym{ne}{NE}{Nash Equilibrium}
\newacronym{hf}{HF}{High Frequency}
\newacronym{noma}{NOMA}{Non-Orthogonal Multiple Access}
\newacronym{sre}{SRE}{Smart Radio Environment}
\newacronym{ris}{RIS}{Reconfigurable Intelligent Surface}
\newacronym{inp}{InP}{Infrastructure Provider}
\newacronym{smf}{SMF}{Slicing Magangement Framework}
\newacronym{nsn}{NSN}{Network Slicing Negotiation}
\newacronym{sms}{SMS}{Slicing MAC Scheduler}
\newacronym{brd}{BRD}{Best Response Dynamics}
\newacronym{dssbr}{DSSBR}{Double Step Smoothed Best Response}
\newacronym{poa}{PoA}{Price of Anarchy}
\newacronym{pos}{PoS}{Price of Stability}
\newacronym{milp}{MILP}{Mixed Integer-Linear Program}
\newacronym{pod}{PoD}{Price of DSSBR}
\newacronym{roc}{ROC}{Radio Overload Control}
\newacronym{ciot}{cIoT}{critical Internet of Things}
\newacronym{embbpr}{eMBB Pr.}{enhanced Mobile BroadBand Premium}
\newacronym{embbbs}{eMBB Bs.}{enhanced Mobile BroadBand Basic}
\newacronym{en}{EN}{Edge Node}
\newacronym{ec}{EC}{Edge Computing}
\newacronym{sp}{SP}{Service Provider}
\newacronym{me}{ME}{Market Equilibrium}
\newacronym{so}{SO}{Social Optimum}
\newacronym{wso}{WSO}{Weighted Social Optimum}
\newacronym{ps}{PS}{Proportional Sharing}
\newacronym{eg}{EG}{Eisenberg-Gale program}
\newacronym{pe}{PE}{Pareto Efficiency}
\newacronym{nsw}{NSW}{Nash Social Welfare}
\newacronym{ef}{EF}{Envy-Freeness}
\newacronym{sub6}{sub6GHz}{Below 6GHz}
\newacronym{ncr}{NCR}{Network-Controlled Repeater}
\newacronym{nlos}{NLoS}{Non-Line of Sight}
\newacronym{src}{SRC}{Smart Radio Connection}
\newacronym{srd}{SRD}{Smart Radio Device}
\newacronym{cs}{CS}{Candidate Site}
\newacronym{tp}{TP}{Test Point}
\newacronym{fov}{FoV}{Field of View}
\newacronym{nrric}{near-RT RIC}{Near Real-time RAN Intelligent Controller}
\newacronym{e2ap}{E2AP}{E2 Application Protocol}
\newacronym{e2sm}{E2SM}{E2 Service Model}
\newacronym{nrtric}{non-RT RIC}{Non-Real-Time Ran Intelligent Controller}
\newacronym{itti}{ITTI}{Inter-task Interface}
\newacronym{bap}{BAP}{Backhaul Adaptation Protocol}
\newacronym{iabest}{IABEST}{Integrated Access and Backhaul Experimental large-Scale Tetbed}
\newacronym{teid}{TEID}{Tunnel Endpoint Identifier}
\newacronym{dlsch}{DL-SCH}{Downlink Shared Channel }
\newacronym{ulsch}{UL-SCH}{Uplink Shared Channel }
\tikzstyle{startstop} = [rectangle, rounded corners, minimum width=2cm, minimum height=0.5cm,text centered, draw=black]
\tikzstyle{io} = [trapezium, trapezium left angle=70, trapezium right angle=110, minimum width=3cm, minimum height=1cm, text centered, draw=black]
\tikzstyle{process} = [rectangle, minimum width=2cm, minimum height=0.5cm, text centered, draw=black, alignb=center]
\tikzstyle{decision} = [ellipse, minimum width=2cm, minimum height=1cm, text centered, draw=black]
\tikzstyle{arrow} = [thick,<->,>=stealth]
\tikzstyle{line} = [thick,>=stealth]
\tikzstyle{darrow} = [thick,<->,>=stealth,dashed]
\tikzstyle{sarrow} = [thick,->,>=stealth]
\tikzstyle{larrow} = [line width=0.1mm,dashdotted,->,>=stealth]
\tikzstyle{llarrow} = [line width=0.1mm,->,>=stealth]

\makeatletter
\def\grd@save@target#1{%
  \def\grd@target{#1}}
\def\grd@save@start#1{%
  \def\grd@start{#1}}
\tikzset{
  grid with coordinates/.style={
    to path={%
      \pgfextra{%
        \edef\grd@@target{(\tikztotarget)}%
        \tikz@scan@one@point\grd@save@target\grd@@target\relax
        \edef\grd@@start{(\tikztostart)}%
        \tikz@scan@one@point\grd@save@start\grd@@start\relax
        \draw[minor help lines] (\tikztostart) grid (\tikztotarget);
        \draw[major help lines] (\tikztostart) grid (\tikztotarget);
        \grd@start
        \pgfmathsetmacro{\grd@xa}{\the\pgf@x/1cm}
        \pgfmathsetmacro{\grd@ya}{\the\pgf@y/1cm}
        \grd@target
        \pgfmathsetmacro{\grd@xb}{\the\pgf@x/1cm}
        \pgfmathsetmacro{\grd@yb}{\the\pgf@y/1cm}
        \pgfmathsetmacro{\grd@xc}{\grd@xa + \pgfkeysvalueof{/tikz/grid with coordinates/major step x}}
        \pgfmathsetmacro{\grd@yc}{\grd@ya + \pgfkeysvalueof{/tikz/grid with coordinates/major step y}}
        \foreach \x in {\grd@xa,\grd@xc,...,\grd@xb}
        \node[anchor=north] at (\x,\grd@ya) {\pgfmathprintnumber{\x}};
        \foreach \y in {\grd@ya,\grd@yc,...,\grd@yb}
        \node[anchor=east] at (\grd@xa,\y) {\pgfmathprintnumber{\y}};
      }
    }
  },
  minor help lines/.style={
    help lines,
    gray,
    line cap =round,
    xstep=\pgfkeysvalueof{/tikz/grid with coordinates/minor step x},
    ystep=\pgfkeysvalueof{/tikz/grid with coordinates/minor step y}
  },
  major help lines/.style={
    help lines,
    line cap =round,
    line width=\pgfkeysvalueof{/tikz/grid with coordinates/major line width},
    xstep=\pgfkeysvalueof{/tikz/grid with coordinates/major step x},
    ystep=\pgfkeysvalueof{/tikz/grid with coordinates/major step y}
  },
  grid with coordinates/.cd,
  minor step x/.initial=.5,
  minor step y/.initial=.2,
  major step x/.initial=1,
  major step y/.initial=1,
  major line width/.initial=1pt,
}
\makeatother

\definecolor{desireRed}{RGB}{230,57,60}%
\definecolor{darkPurple}{RGB}{59,31,43}%
\definecolor{springGreen}{RGB}{37,223,145}%
\definecolor{queenBlue}{RGB}{69,123,157}%
\definecolor{spaceCadet}{RGB}{29,53,87}%

\usepackage{dblfloatfix}

\begin{document}

\title{Toward Open Integrated Access and Backhaul\\with O-RAN}

\author{
    \IEEEauthorblockN{Eugenio Moro\IEEEauthorrefmark{1}, Gabriele Gemmi\IEEEauthorrefmark{2}, Michele Polese\IEEEauthorrefmark{3}, Leonardo Maccari\IEEEauthorrefmark{2}, Antonio Capone\IEEEauthorrefmark{1}, Tommaso Melodia\IEEEauthorrefmark{3}}\\
    \IEEEauthorblockA{\IEEEauthorrefmark{1}Department of Electronics, Information and Bioengineering,\\
    Polytechnic University of Milan, Italy
    \\\{name.surname\}@polimi.it}
    \IEEEauthorblockA{\IEEEauthorrefmark{2}Department of Environmental Sciences, Informatics and Statistics,\\
    Ca' Foscari University of Venice, Italy.
    \\\{name.surname\}@unive.it}
    \IEEEauthorblockA{\IEEEauthorrefmark{3}Institute for the Wireless Internet of Things,\\
    Northeastern University, Boston, MA, U.S.A.
    \\\{n.surname\}@northeastern.edu}
\thanks{This work was partially supported by NGIAtlantic.eu project within the EUHorizon 2020 programme under Grant No. 871582, by the U.S. National Science Foundation under Grant CNS-1925601, and by OUSD(R\&E) through Army Research Laboratory Cooperative Agreement Number W911NF-19-2-0221. The views and conclusions contained in this document are those of the authors and should not be interpreted as representing the official policies, either expressed or implied, of the Army Research Laboratory or the U.S. Government. The U.S. Government is authorized to reproduce and distribute reprints for Government purposes notwithstanding any copyright notation herein.}
}

\maketitle

\begin{abstract}
\Gls{mmwave} communications has been recently standardized for use in the fifth generation (5G) of cellular networks, fulfilling the promise of multi-gigabit mobile throughput of current and future mobile radio network generations. In this context, the network densification required to overcome the difficult \gls{mmwave} propagation will result in increased deployment costs. \gls{iab} has been proposed as an effective mean of reducing densification costs by deploying a wireless mesh network of base stations, where backhaul and access transmissions share the same radio technology. However, IAB requires sophisticated control mechanisms to operate efficiently and address the increased complexity. The Open \gls{ran} paradigm represents the ideal enabler of RAN intelligent control, but its current specifications are not compatible with IAB. In this work, we discuss the challenges of integrating IAB into the Open RAN ecosystem, detailing the required architectural extensions that will enable dynamic control of 5G IAB networks. We implement the proposed integrated architecture into the first publicly-available Open-RAN-enabled experimental framework, which allows prototyping and testing Open-RAN-based solutions over end-to-end 5G IAB networks. Finally, we validate the framework with both ideal and realistic deployment scenarios exploiting the large-scale testing capabilities of publicly available experimental platforms. 
\end{abstract}

 
\glsresetall

\section{Introduction}
\label{sec:introduction}
\gls{ran} densification is a key technique to boost the coverage and performance metrics of current and future generations of mobile radio networks~\cite{dang2020should}.
However, these ultra-dense deployments come with increased costs and complexity for provisioning wired backhaul to each base station~\cite{bushan2014network}. To address this, the \gls{3gpp} has introduced \gls{iab} in its Release 16 for NR~\cite{takao2020}. 
With \gls{iab}, the backhaul traffic is multiplexed on the air interface together with regular \glspl{ue} access traffic. This effectively creates a wireless mesh network of \glspl{bs} where only a few require an expensive wired connection to the \gls{cn} (i.e., the IAB-Donors). Hence the cost-reduction potential through wireless relays (i.e., the IAB-Nodes)~\cite{polese2020iab}. Additionally, \gls{iab} is especially relevant for \gls{mmwave}-based radio access, where inexpensive network densification is a fundamental necessity~\cite{feng2017mmwave}.

While the standardization process has reached a sufficient maturity level, the open challenges brought about by integrating access and backhaul are still open. Consequently, \gls{iab} offers optimization opportunities at all layers of communication abstraction. 
At the lowest levels, specialized \gls{iab}-aware techniques are required to ensure a fair and effective resource allocation among \glspl{ue} and \glspl{mt}~\cite{zhang2020sched,zhang2021resource}. 
At the same time, backhaul and access transmission multiplexing must be managed to minimize interference~\cite{yu2023coordinated}. 
Furthermore, adaptive topology reconfiguration mechanisms must be provisioned to maintain resiliency against link failures, traffic unbalances and anomalous user distribution~\cite{ranjan2021cellselection}. Overall, these sophisticated management procedures require control primitives that go beyond what has been specified by \gls{3gpp}.

\glsunset{oran}
The unprecedented paradigm shift brought about by the \gls{oran} architecture, developed by the O-RAN Alliance, promises to enable programmatic control of \gls{ran} components through open interfaces and centralized control loops~\cite{polese2022understanding}. 
As such, it is the ideal candidate to unlock the potential optimization and management gains awaiting in \gls{iab}. 
However, the current \gls{oran} architecture is tailored to traditional \gls{ran} deployments, and an extension to enable \gls{iab} control is required. 
The first contribution of this work resides in a discussion on how the O-RAN architecture, interfaces, and control loops can be extended to IAB scenarios, with the ultimate goal of allowing large-scale, data-driven control and management of \gls{5g} \gls{iab} networks.

Additionally, to foster prototyping and testing with \gls{iab} and \gls{oran}, we propose a comprehensive framework where researchers can easily deploy an end-to-end \gls{oran}-enabled IAB network with \gls{ota} and hardware-in-the-loop emulation capabilities. 
In line with \gls{oran} core concepts, our framework is designed to be open, accessible and flexible by leveraging on open-source software and \gls{cots} hardware. 
The framework builds on IABEST, the first large-scale accessible and open \gls{iab} testbed presented in~\cite{moro2022iabest}. 
This testbed has been enriched to produce a complete \gls{oran} \gls{iab} experimental solution, effectively replicating the proposed \gls{oran}-\gls{iab} integrated architecture. 
In particular, IAB-Donors and IAB-Nodes have been equipped with custom-developed agents for the so-called E2 and O1 standard interfaces. 
These additions enable the controllers introduced by the \gls{oran} architecture to manage IAB-Nodes, effectively representing the first publicly available \gls{oran}-enabled \gls{iab} prototyping and testing solution.

To further facilitate experimental research activities, we have packaged and integrated the entire framework into OpenRAN Gym, a publicly-available research platform for data-driven O-RAN experimentation at scale~\cite{bonati2023openrangympawr}. 
Through OpenRAN Gym, researchers can swiftly deploy and test the proposed framework over large-scale and publicly available hardware experimental platforms, such as the PAWR testbeds and Colosseum~\cite{manu2018pawr,bonati2021colosseum}.  
Notably, we showcase how Colosseum can be leveraged for large-scale \gls{iab} testing through hardware-in-the-loop channel emulation to create sophisticated deployment scenarios. A tutorial on how to deploy an O-RAN-driven IAB network, together with the source code of all the framework, is available on the OpenRAN Gym website.\footnote{\url{https://openrangym.com/tutorials/iab-tutorial}}
Finally, we use Colosseum to validate the proposed framework numerically. In particular, we test the attainable performance in a controlled radio scenario and in a more realistic deployment in which we reconstruct a part of Florence, Italy. 

The remainder of this paper is organized as follows. 
Section~\ref{sec:architecture} analyses the challenges of extending \gls{oran} to \gls{5g} \gls{iab} networks. 
Section~\ref{sec:experimental} contains a description of the proposed frameworks, focusing on the \gls{oran} extensions that have been included in~\cite{moro2022iabest}. 
Section~\ref{sec:results} contains the results of the experiments we performed to validate our framework by exploiting the large-scale testing capabilities of Colosseum. 
Finally, Section~\ref{sec:conclusions} concludes the paper and discusses future extensions.

\begin{figure*}[ht]
    \centering
    \includegraphics[width=.9\textwidth]{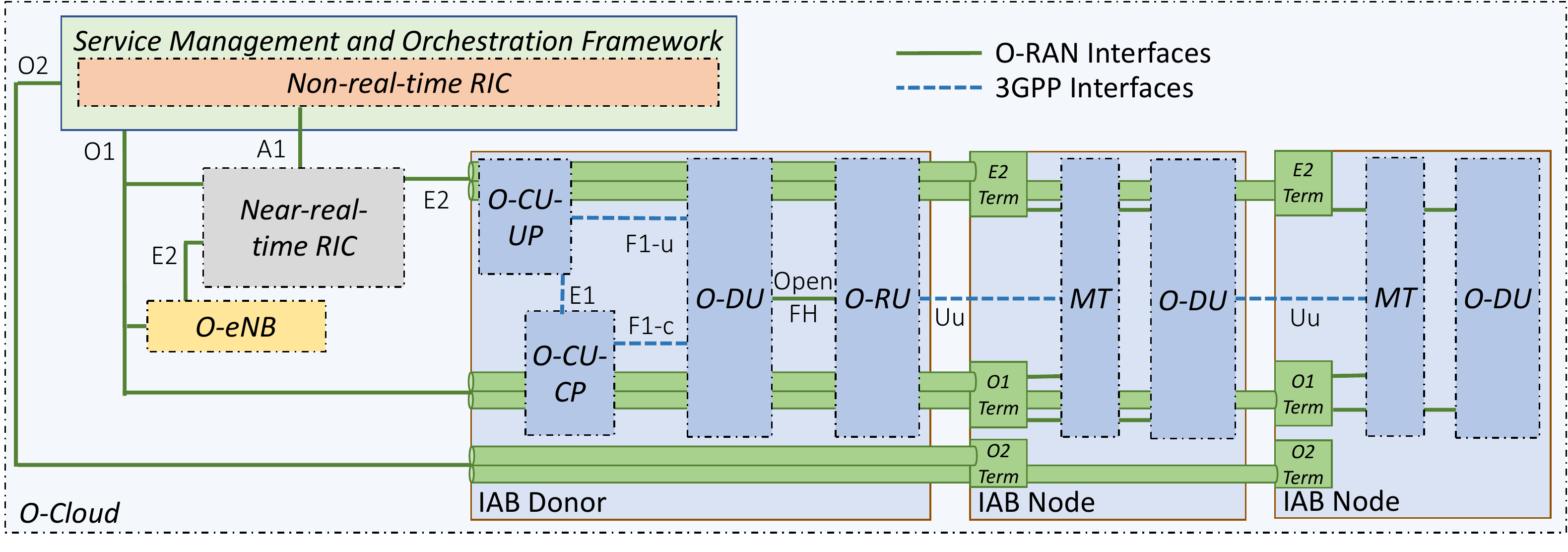}
    \caption{IAB and O-RAN integrated architectures.}
    \label{fig:iab-o-ran}
\end{figure*}

\section{Integrating  \gls{iab} in  Open \gls{ran}}
\label{sec:architecture}

As discussed in Section~\ref{sec:introduction}, \gls{iab} represents a scalable solution to the need for backhaul in ultra-dense 5G and 6G deployments. At the same time, however, the wireless backhaul introduces additional complexity to the network deployments: new parameters and configurations that need to be tuned---and possibly, adapted dynamically---to get the best performance out of the network and to seamlessly adjust to updated conditions in the scenario and in the equipment status. 
For example, it is possible to optimize the \gls{iab} network performance by properly selecting the connectivity of \gls{iab}-Nodes to their parents~\cite{ranjan2021cellselection}, or by appropriately allocating resources to backhaul and access flows sharing the same air interface~\cite{zhang2020sched}.

As for traditional \gls{ran} deployments with fiber-based backhaul~\cite{niknam2022intelligent}, there is a case to be made for providing \gls{iab} \gls{ran} equipment with primitives for flexible, dynamic, data-driven programmatic control. 
This requires providing endpoints to expose telemetry, measurements, and analytics from \gls{iab}-Nodes, as well as parameters and control knobs to enable the optimization. So far, the Open \gls{ran} paradigm has been successfully applied to non-\gls{iab} networks to achieve the same goals, thanks to interfaces that give access to \gls{3gpp} \glspl{kpm} and control parameters in the \gls{ran} nodes~\cite{lee2020hosting,brik2022deep}. 
The Open \gls{ran} vision, which is being developed into technical specifications by the O-RAN Alliance, includes controllers that run custom control loops, i.e., the \glspl{ric}. 
The O-RAN Alliance has defined control loops and related \glspl{ric} that can operate at a time scale of 10 ms to 1 s (i.e., \emph{near-real-time}) or more than 1 s (i.e., \emph{non-real-time})~\cite{3gppORAN}. 
The near-real-time, or near-RT, \gls{ric} is connected to the \gls{ran} nodes through the E2 interface, while the non-real-time \gls{ric}, which is part of the network \gls{smo}, interacts with the \gls{ran} through the O1 interface, as shown in the left part of Figure~\ref{fig:iab-o-ran}. Other interfaces from the non-RT \gls{ric}/\gls{smo} include A1 to the near-RT \gls{ric}, for policy guidance and \gls{ai}/\gls{ml} model management, and the O2 interface to the O-Cloud, which is an abstraction of the virtualization infrastructure that can support the deployment of O-RAN functions. 
The use of standard interfaces makes it possible to run even third-party applications in the controllers, the so-called \textit{xApps} and \textit{rApps} for the \gls{nrric} and \gls{nrtric}, respectively.

The \gls{3gpp} already provides control and adaptation capabilities through the \gls{iab} \gls{bap} layer, the F1 interface, and the \gls{rrc} layer across the \gls{iab}-Donor \gls{cu} and the \gls{iab}-Node \gls{du}. How and when control and adaptation of such configurations could be performed, however, is left to the vendor implementation. 
This is where an extension of the O-RAN architecture to \gls{iab} networks can play a role, exposing \gls{iab}-Donor and \gls{iab}-Node functions to the \glspl{ric}. 
These can leverage a centralized point of view on the \gls{ran} and a wealth of analytics and information usually unavailable in the individual \gls{iab}-Donors and Nodes. For \gls{iab}, this could translate into effective multi-donor coordination with reduced interference and agile topology adaptation across different \gls{iab}-Donor domains, and dynamic resource allocation with---for example---data-driven proactive congestion identification and resolution across access and backhaul links.

\subsection{Extensions to Open \gls{ran}}
Extending the O-RAN architecture and interfaces to \gls{iab} deployments, however, presents some design and architectural challenges. 
Primarily, supporting O-RAN interfaces in \gls{iab}-Nodes means either (i) terminating the interfaces at the \gls{iab}-Donor; or (ii) transporting their data over the wireless backhaul. 
The first option is simpler, does not require architectural updates, but at the same time limits the control and reconfiguration to what is available in the \gls{iab}-Donor, without insight on the \gls{iab}-Nodes. 
The second option, instead, provides more granular access at the cost of additional complexity and tunneling of data over the wireless backhaul. 

The \gls{3gpp} already foresees performing \gls{smo}-like operations through the wireless backhaul interface~\cite{3gpp_gnb_split}. Therefore, in this paper and in the architecture described in Figure~\ref{fig:iab-o-ran} we consider the second option, which would provide tighter and more effective integration between O-RAN and \gls{iab} deployments.
In general, the tunneling can be performed by encapsulating the O-RAN interfaces payloads into dedicated bearers.
Note that this requires some interaction between functions of the control plane of the network and the transport in the user plane, e.g., through a dedicated \gls{pdu} session between a local \gls{upf} in the IAB-Donor and in the \gls{iab}-Node \gls{mt}. 
Then, a local interface termination can be installed in the \gls{iab}-Node, as it would in a traditional, fiber-equipped \gls{ran} node. 
The O-RAN traffic, in this case, would be multiplexed with user data on the wireless backhaul resources, and it needs to be properly prioritized to achieve the control goals while not harming users' performance or creating congestion. 
\begin{figure*}[ht]
  \centering
  \includegraphics[width=0.9\textwidth]{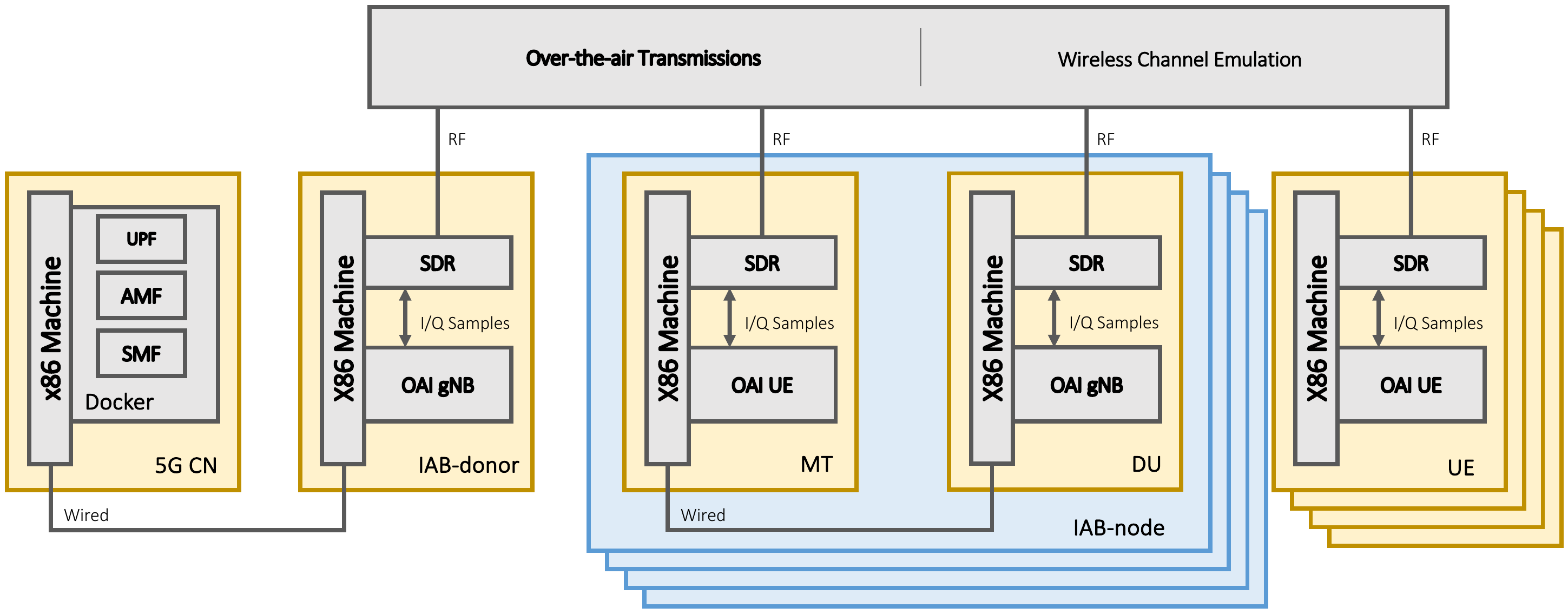}
  \setlength\belowcaptionskip{-.3cm}
  \setlength\abovecaptionskip{0.2cm}
  \caption{Overview of the RAN architecture deployed over white-box hardware.}
  \label{fig:iab_arch}
\end{figure*}

\textbf{E2 extension for \gls{iab}.} The extension of the E2 interface likely requires one or multiple new, dedicated \glspl{e2sm}. 
The \gls{e2sm} represents the semantic of the E2 interface, i.e., the \gls{ran} function with which an xApp in the near-RT \gls{ric} interacts. 
For \gls{iab}, an extension of \gls{e2sm} \gls{kpm}~\cite{oran_kpm} can be used to expose performance metrics related to the \gls{mt}, besides the \gls{du}. Another near-real-time control target over E2 can include, for example, resource partitioning between backhaul and access traffic, or dynamic \gls{tdd} slot configuration to adapt to varying traffic on the access and backhaul.

\textbf{O1 extension for \gls{iab}.} The O1 interface would connect the \gls{smo} to the \gls{iab}-Node, e.g., to perform maintenance and updates of the components (\gls{mt} and \gls{du}) of the \gls{iab}-Node. 
Compared to E2 near-real-time control, the O1 interface would run control loops at 1 s or more. Thus its traffic can be transported with lower priority than the E2 traffic. 
This makes a case for dedicated bearers and tunnels on the backhaul interface for \emph{each} of the O-RAN interfaces. 

\textbf{O2 extension for \gls{iab}.} This interface can be used to integrate the \gls{iab}-Nodes as resources in the O-Cloud. 
Compared to traditional virtualization infrastructure for the O-Cloud, the \gls{iab}-Nodes are available---and reachable over O2---only when a session is established from one \gls{iab}-Donor to the \gls{iab}-Node itself.

\section{An Experimental Framework for\\\gls{iab} and \gls{oran}}
\label{sec:experimental}

Our proposed experimental framework packages the entire software chain required to run the \gls{oran}-enabled \gls{iab} network described in Section~\ref{sec:architecture} in a multi-layer architecture. 
At the hardware level, our framework does not present any specific requirement. 
Indeed, every software component can run on \gls{cots} hardware like generic x86 machines and USRP \gls{sdr}. 
On the other hand, some software components are customized or designed from scratch to reproduce and support a \gls{5g} \gls{iab} network. 
In particular, we have adapted \gls{oai}, an open source \gls{5g} \gls{ran} framework~\cite{kaltenberger2020openairinterface}, to implement \gls{iab}-Donors, \gls{iab}-Nodes, and \gls{iab}-capable core functions. 
Additionally, we have integrated agents for the E2 and O1 interfaces in the IAB-Donor and IAB-Node, effectively implementing the architectural integration proposed in Section~\ref{sec:architecture}.
These interfaces are used by the non-real-time and real-time \glspl{ric} packaged in our framework to control all the components of the deployed IAB network. 
We now describe the aforementioned components, separating them into the \gls{ran} and \gls{oran} domains.

\subsection{RAN and Core Network Components}
Figure~\ref{fig:iab_arch} represents an overview of the radio access functional components that enable end-to-end communication in our framework.
In particular, we provide the following: a minimal yet functional deployment of \gls{5g} \gls{cn} functions, software-defined \gls{iab}-Nodes and IAB-Donors and software-defined \glspl{ue}.

\textbf{IAB-Nodes and IAB-Donors.}
According to \gls{3gpp} specifications~\cite{polese2020iab}, an IAB-Donor hosts a \gls{cu} and multiple \glspl{du}. Similarly, IAB-Node is split into a \gls{du} and an \gls{mt}.
Functionally, these have the task of enabling downstream and upstream connectivity, respectively. At the time of writing, \gls{oai}'s main branch implementation of the \gls{cu}/\gls{du} functional split does not support multiple \glspl{du} connected to a single \gls{cu}~\cite{oaif1}.
This limitation is incompatible with the \gls{iab} architecture. Consequently, we employ a full \gls{oai} \gls{gnb} in place of both \gls{cu} and \gls{du}. In other words, the IAB-Nodes and IAB-Donors in our framework do not follow \gls{3gpp} split 2. Instead, these components are deployed as monolithic \glspl{gnb}.
As for the \gls{mt}, an open-source implementation is currently unavailable. However, this component is functionally equivalent to a \gls{ue}, as it connects to upstream nodes using the same resources and protocols. Consequently, we have selected \gls{oai}'s software-defined \gls{ue} to act as \glspl{mt} in the proposed framework. This results in a system where a single \gls{iab}-Node is made up of two concurrently running instances: an \gls{oai} \gls{gnb}---acting as a \gls{du}---and an \gls{oai} \gls{ue}---acting as a \gls{mt}. 
In the resulting architecture, \gls{iab}-Nodes are naturally deployed over two separate machines, hosting the \gls{gnb} and the \gls{ue}, and connected out-of-band as it is shown in Figure~\ref{fig:iab_arch}. Alternatively, the two software components can run on a single x86 machine, provided that sufficient computing power is available.
While this architecture does not require any particular modification to \gls{oai}'s implementations, we have added a signaling functionality through which the \gls{iab}-Nodes or \gls{iab}-Donors can discern connected \glspl{mt} from connected \glspl{ue}. This has been achieved through proper manipulation of the UE Capability messages. 
Such information can be exploited to enable \gls{iab}-aware optimization solutions in the \gls{gnb}. 

\textbf{Core Network Functions.}
A minimal set of \gls{5g} \gls{cn} functions have been included in our framework: \gls{nrf}, \gls{amf}, \gls{smf} and \acrfull{upf}, all based on the \gls{oai} 5G core implementation. All these functions run as containers on a single x86 machine, as shown in Figure~\ref{fig:iab_arch}. Due to the selected \gls{iab} system design, the \gls{upf} required modifications to enable \gls{iab} operations.
As previously mentioned, \glspl{ue} acts as \glspl{mt} in \gls{iab}-Nodes, connecting to upstream nodes. The established \gls{gtp} tunnels are then used to provide direct connectivity between the \gls{du} of the node and the \gls{cn} functions. In other words, \gls{mt}-acting \glspl{ue} relay the backhaul traffic of the \gls{iab}-Nodes.
However, \gls{oai}'s \gls{upf} implementation lacks support for the required forwarding capability,\footnote{To the best of the authors' knowledge, there is no available open source implementation that supports this operating mode.} as any packet whose destination is not a \gls{ue} is dropped.
Therefore, we have implemented a minimal version of framed routing~\cite{3gpp.29.244} in \gls{oai} \gls{upf}, enabling \glspl{ue} to act as intermediate nodes. 

\textbf{User Equipment}
From the perspective of the \gls{ue}, an \gls{iab} network deployed using the components described above is entirely standard-compliant. As such, both software-defined \glspl{ue} (as shown in Figure~\ref{fig:iab_arch}) and \gls{cots} \glspl{ue} can be used in the proposed framework. 

\subsection{O-RAN Components}
As mentioned in Section~\ref{sec:architecture}, \gls{oran} defines a set of standardized and open interfaces with which the \gls{ran} exposes data collection and control primitives to the \glspl{ric}. 
In the proposed framework, we have enabled IAB-Nodes and IAB-Donors to be \gls{oran}-compatible by integrating software agents for the E2 and O1 interfaces into the codebase of \gls{oai}. Furthermore, our framework comprises a \gls{nrric} and a \gls{nrtric}.

\textbf{E2 interface integration.}
The E2 interface is functionally split into two protocols: E2AP---tasked with establishing a connection with the \gls{nrric}---and E2SM---which implements specific monitoring and control functionalities, namely \glspl{sm}, as discussed in Section~\ref{sec:architecture}. 
In the software implementation we provide, E2AP has been adapted from \gls{oran} Alliance Software Community reference implementation and, as such, it is entirely compliant with \gls{oran}.
On the other hand, the \glspl{sm} provided by the \gls{oran} alliance are defined using ASN.1: a powerful production-ready abstract description language which is, however, cumbersome and challenging to use in the fast-paced research and development environments targeted by our framework.
In light of this, we employ custom \gls{sm} that are defined through Protocol Buffers (protobuf)---an abstract definition language that is easier to handle and allows for fast prototyping and testing, facilitating the development of \gls{iab}-aware control solutions. 
Since the E2 interface is such that the E2SM messages are encoded and decoded only in the \gls{ran} and xApp, the custom \gls{sm} definitions are transparent to the RIC, allowing our proposed solution to retain generic \gls{oran} compliance. At the time of this writing, we have implemented a set of protobuf messages that can be used to reproduce both the \gls{kpm} and \gls{rc} \glspl{sm}~\cite{polese2022understanding}. These can be used to develop data collection and control xApps, respectively. 

\begin{figure*}[ht]
    \centering
    \subcaptionbox{Linear IAB topology.\label{fig:linear_topo}}{
        \raisebox{20pt}{
        \includegraphics[width=0.28\linewidth]{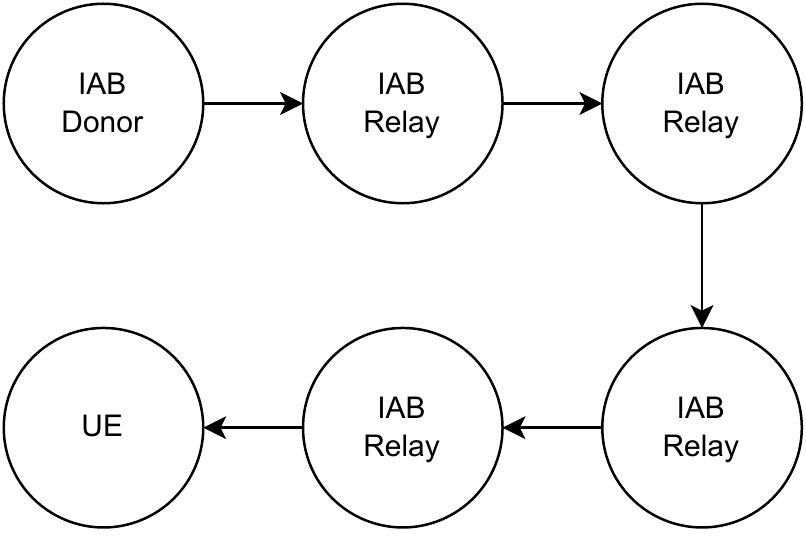}}
        }
    \hfill
    \subcaptionbox{Throughput measurements for the linear topology.\label{fig:tp_linear}}{
        \begin{tikzpicture}
    \pgfplotsset{every tick label/.append style={font=\scriptsize}}
    \begin{axis}[
        width=.3\linewidth, %
        grid=major, %
        grid style={dashed,gray!50}, %
        xlabel=Hops, %
        xtick = {1, 2, 3, 4, 5},
        ytick = {10, 20, 30, 40, 50, 60},
        ylabel={Throughput [Mbps]},
        ylabel style ={font=\footnotesize},
        xlabel style ={font=\footnotesize},
        error bars/y dir=both, %
        error bars/y explicit,  %
        legend style={at={(0.99,0.99)},anchor=north east,font=\footnotesize}, %
      ]
      \addplot[black,mark=o] table[x=hops,y=thr,y error=std,col sep=comma] {data/linear/dl.csv}; 
      \addlegendentry{Downlink}
      \addplot[black,dashed,mark=star] table[x=hops,y=thr,y error=std,col sep=comma] {data/linear/ul.csv}; 
      \addlegendentry{Uplink}
    \end{axis}
\end{tikzpicture}
    }
    \hfill
    \subcaptionbox{\Gls{rtt} measures for the linear topology.\label{fig:rtt_linear}}{
        \begin{tikzpicture}
    \pgfplotsset{every tick label/.append style={font=\scriptsize}}
    \begin{axis}[
        width=.3\linewidth, %
        grid=major, %
        grid style={dashed,gray!50}, %
        xlabel=Hops, %
        ylabel={RTT [ms]},
        xtick = {1, 2, 3, 4, 5},
        ytick = {20, 40, 60, 80},
        color=black,
        ylabel style ={font=\footnotesize},
        xlabel style ={font=\footnotesize},
        error bars/y dir=both, %
        error bars/y explicit,  %
        legend style={at={(0.5,-0.2)},anchor=north,font=\footnotesize}, %
      ]
      \addplot[black,mark=o] table[x=hops,y=delay,y error=std,col sep=comma] {data/linear/rtt.csv}; 
    \end{axis}
\end{tikzpicture}
    }
    \caption{Topology and results for the linear chain.}
\end{figure*}

\textbf{O1 interface integration.}
In order to properly manage all the different aspects of networked elements, the O1 interface defines various \gls{mns}, which can be used either from the managed entities (the \glspl{gnb}) to report information back to the \gls{ric} or from the managing entity (the \gls{smo} and the rApps running on it) to deploy configurations changes, transfer files or update the software on the managed entities~\cite{oranwg10,polese2022understanding}. 
Among all the different \gls{mns}, we have focused our contribution on implementing the Heartbeat \gls{mns}, which periodically transmits heartbeats; the Fault Supervision \gls{mns}, which reports errors and events; and the Performance Assurance \gls{mns}, which streams performance data. 
Those \gls{mns} have been integrated into the \gls{oai} codebase by implementing a scheduler that, running on a dedicated thread, periodically sends \gls{ves} notifications in JSON format over HTTP. 
This format and communication protocol has been chosen among the different options defined in the standard, as it is widely known and easily extendable by other researchers. 
As of now, our implementation reports performance metrics, such as the throughput and information on the channel quality between \gls{iab}-Nodes, and failure events, such as \gls{rrc} or \gls{ulsch} failures, which can be used in rApps to monitor and optimize the backhaul network.
Provisioning \gls{mns}, which can be used by the rApps to deploy configuration changes (e.g., topology optimizations), have not been implemented by following the O1 specifications, as it would have needed major reworks in the \gls{oai} codebase. 
Instead, we have taken advantage of \textit{IAB-Manager}, a software component we developed to orchestrate \gls{iab} experiments, as discussed next.

\textbf{IAB-Manager.}
IAB networks are generally expected to include several IAB-Nodes, and the proposed framework can scale to such numbers. However, managing experiments with tens or more \gls{ran} components can take time and effort. Indeed, each component is potentially hosted by a dedicated machine, and setting up an \gls{iab} deployment requires each one to be activated and configured according to a sequence that starts from the \gls{cn} functions and ends with the terminal \gls{iab}-Nodes. 
To facilitate experimenting at such a large scale, we have developed \textit{IAB-Manager}~\cite{moro2022iabest}: a software component that can automate the \gls{iab} network deployment and testing through a command line interface and an \gls{api}. 
In particular, \textit{IAB-Manager} is a single entrypoint for controlling the entire experiment: network components and radio environment setup (in case of wireless channel emulation), topology and routing management and reconfiguration, automated testing, and result collection. 
From a functional perspective, the manager connects to the machines involved in the experimentation and configures them according to the assigned roles. In particular, once the user specifies the roles, the manager sequentially activates each network component until the final deployment is ready for experimentation, greatly simplifying the setup phase. Additionally, as previously mentioned, \textit{IAB-Manager} executes the network configuration changes mandated by the rApps. 

\textbf{RAN Intelligent Controllers.}
The proposed framework packages a \gls{nrric} and a \gls{nrtric}. Both are compliant with the standard and based on \gls{oran} Software Community reference implementations.

\section{Validation and Results}\label{sec:results}

\begin{table}
    \centering
    \begin{tabular}{ll}
        \toprule
        Parameter & Value \\
        \midrule
        Area Size for realistic deployment & 0.627 km$^2$ \\
        gNB Density & 45 gNB/km$^2$ \\
        \gls{iab}-donors/ \gls{iab}-nodes ratio & 1/10 \\
        Emulated center frequency & 28 GHz \\
        Bandwidth & 40 MHz \\
        Scheduler & 7 2 1 \\
        Subcarrier Spacing & 30khz \\
        Colosseum Base loss & 50 dB \\
        3GPP Channel Model & Urban Micro \\
        MIMO layers & 1 \\      
        \bottomrule
    \end{tabular}
    \vspace{1em}
    \caption{Table of System Settings} \label{tab:simulation_params}
\end{table}

This section focuses on validating our proposed framework from an experimental perspective. 
In particular, we are interested in giving an initial characterization of some fundamental \glspl{kpi} of the deployments allowed by our \gls{iab} framework while validating its correct functioning.

While the openness and flexibility of the software components are such that the framework can run on generic hardware, we chose to base our validation campaign on Colosseum~\cite{bonati2021colosseum}. 
The Colosseum testbed is a publicly available large-scale testing platform with hardware-in-the-loop capabilities. 
It comprises 128 \glspl{srn}, each composed of a powerful x86 computing node and an USRP X310 \gls{sdr}. 
All the components in the proposed framework can be easily deployed on \glspl{srn}. 
Every \gls{srn} radio is interconnected by an FPGA mesh that emulates arbitrary radio channels defined through tapered delay models. 
With the capability of emulating complex scenarios of tens of entities, Colosseum makes it possible to deploy large IAB networks over complex propagation scenarios. 
As such, it represents an ideal validation platform for our framework. 
Furthermore, Colosseum is open to the research community, and the validation tools are made available, allowing interested parties to start experimenting with a minimal initial effort. 

\subsection{Experiments with a linear chain}

We start by evaluating the performance of an IAB network deployed in a tightly controlled scenario. 
To this end, we consider a 5-hop linear topology, as shown in Figure~\ref{fig:linear_topo}. 
As detailed in Section~\ref{sec:architecture}, each IAB-Node comprises an MT and a DU, bringing this experiment's overall radio node count to 10. 
In order to characterize the upper-bound performance of the proposed framework, we employ an ideal propagation scenario. 
Through properly manipulating Colosseum's channel emulator, a 0~dB pathloss model is selected for nodes connected in the linear topology, and an infinite pathloss is set for all the other channels, effectively suppressing any possible interference. 
In other words, this radio scenario is equivalent to connecting the SDRs with coaxial cables.\footnote{} 
Transmissions occur on band n78 with 106 \glspl{prb} available, for a total of 40MHz bandwidth.

\setcounter{figure}{4}    
\begin{figure*}[b]
    \centering
    \subcaptionbox{Downlink.\label{fig:florence_meas_dl}}{
        \begin{tikzpicture}
    \pgfplotsset{every tick label/.append style={font=\scriptsize}}
    \begin{axis}[
        width=.3\linewidth,
        boxplot/draw direction = y,            
        enlarge y limits,
        ymajorgrids,
        yminorgrids,
        grid style={dashed,gray!40}, %
        xtick = {1, 2, 3},
        ytick = {0,10,20,30,40},
        xticklabel style = {align=center, font=\scriptsize},
        yticklabel style = {align=center, font=\scriptsize},
        xticklabels = {1, 2, 3},
        ylabel style ={font=\footnotesize},
        xlabel style ={font=\footnotesize},
        ylabel = {Throughput [Mbps]},
        xlabel = Hops,
    ]
        \foreach \n in {0,...,2} {
            \addplot+[boxplot, draw=black] table[y index=\n, col sep=comma] {data/firenze_45/dl.csv};
        }
    \end{axis}
\end{tikzpicture}
    }
        \hfill
    \subcaptionbox{Uplink.\label{fig:florence_meas_ul}}{
        \begin{tikzpicture}
    \begin{axis}[
        width=.3\linewidth,
        boxplot/draw direction = y,            
        enlarge y limits,
        ymajorgrids,
        yminorgrids,
        xmajorgrids=false,
        grid style={dashed,gray!40}, %
        xtick = {1, 2, 3},
        ytick = {0,5,10,15,20},
        xticklabel style = {align=center, font=\scriptsize},
        yticklabel style = {align=center, font=\scriptsize},
        xticklabels = {1, 2, 3},
        ylabel style ={font=\footnotesize},
        xlabel style ={font=\footnotesize},
        ylabel = {Throughput [Mbps]},
        xlabel = Hops,
    ]
        \foreach \n in {0,...,2} {
            \addplot+[boxplot, draw=black] table[y index=\n, col sep=comma] {data/firenze_45/ul.csv};
        }
    \end{axis}
\end{tikzpicture}
    }
        \hfill
    \subcaptionbox{Round Trip Time.\label{fig:florence_meas_rtt}}{
        \begin{tikzpicture}
    \begin{axis}[
        width=.3\linewidth,
        boxplot/draw direction = y,            
        enlarge y limits,
        ymajorgrids,
        yminorgrids,
        xmajorgrids=false,
        grid style={dashed,gray!40}, %
        xtick = {1, 2, 3},
        ytick = {0,20,40,60,80},
        xticklabel style = {align=center, font=\scriptsize},
        yticklabel style = {align=center, font=\scriptsize},
        xticklabels = {1, 2, 3},
        ylabel = {RTT [ms]},
        ylabel style ={font=\footnotesize},
        xlabel style ={font=\footnotesize},
        xlabel = Hops,
    ]
        \foreach \n in {0,...,2} {
            \addplot+[boxplot, draw=black] table[y index=\n, col sep=comma] {data/firenze_45/rtt.csv};
        }
    \end{axis}
\end{tikzpicture}
    }
    \caption{Measurements for the realistic scenario.}
    \label{fig:florence_meas}
\end{figure*}

Figure~\ref{fig:tp_linear} shows the downlink and uplink TCP throughput against the number of hops, as measured between the core network and the specific MT/UE. 
The first-hop values of 47~Mbps in DL and 21~Mbps in UL represent the maximum throughput attainable in the testing settings. 
This upper bound is far from the theoretical maximum allowed by the available bandwidth. It is limited by several factors that depend on the experimental platform, OAI software implementation, and system design. 
Most notably, the strongest detractor to the final throughput performance is given by the OAI implementation of the software-defined UE, which is employed to build the \gls{mt}.
In particular, the OAI UE is inefficient in reception and transmission, thus becoming a bottleneck for the entire communication chain. Efforts are ongoing to improve the performance and stability of this software framework.
Furthermore, the frameworks' system design is such that each IP packet is encapsulated into as many \gls{gtp} packets as the number of hops. 
This increased overhead can cause packet fragmentation with a further negative impact on the overall performance. Furthermore, even if the emulated channel is set to a 0~dB~pathloss, Colosseum's architecture includes an unavoidable base loss of 50~dB~\cite{villa2022} due to characteristics of the hardware architecture. This, together with the aforementioned inefficiencies, make such that packet drops and subsequent retransmissions happen also in this ideal scenario. 

As the number of hops increase, the downlink throughput experiences a sharp decrease before stabilizing on a per-hop loss of around 6~Mbps. The notable throughput loss experienced at the second hop can be explained by observing the standard deviation of the throughput, represented by the whiskers in Figure~\ref{fig:tp_linear}. 
This value is at its maximum for the first hop, suggesting that the first radio link is unstable due to the RX pipeline of the MT being overwhelmed. 
This substantial variability is caused by packet loss and retransmissions and internal buffer overflow, which negatively affect the performance of the second hop, as it is noticeable in the numerical results. 
At the same time, the second hop's throughput standard deviation is lower, as the decreased traffic volume causes less drops in the involved MTs. 
This stabilizing effect propagates down the topology, as both the decreasing standard deviation and the linear per-hop loss testify. 
On the other hand, the uplink throughput is relatively stable and close to the upper bound, even at the fourth hop. 
This is because the limited \gls{oai} \gls{ue} performance and BS scheduling process limits the uplink traffic volume, and the \glspl{gnb} are far from being overwhelmed. 
On the other hand, since the uplink throughput does not significantly decrease from the maximum, the UE's congestion level remains relatively stable and high, as proven by the constant standard deviation values.

\setcounter{figure}{3}    
\begin{figure}[ht]
    \centering
    \includegraphics[width=\linewidth]{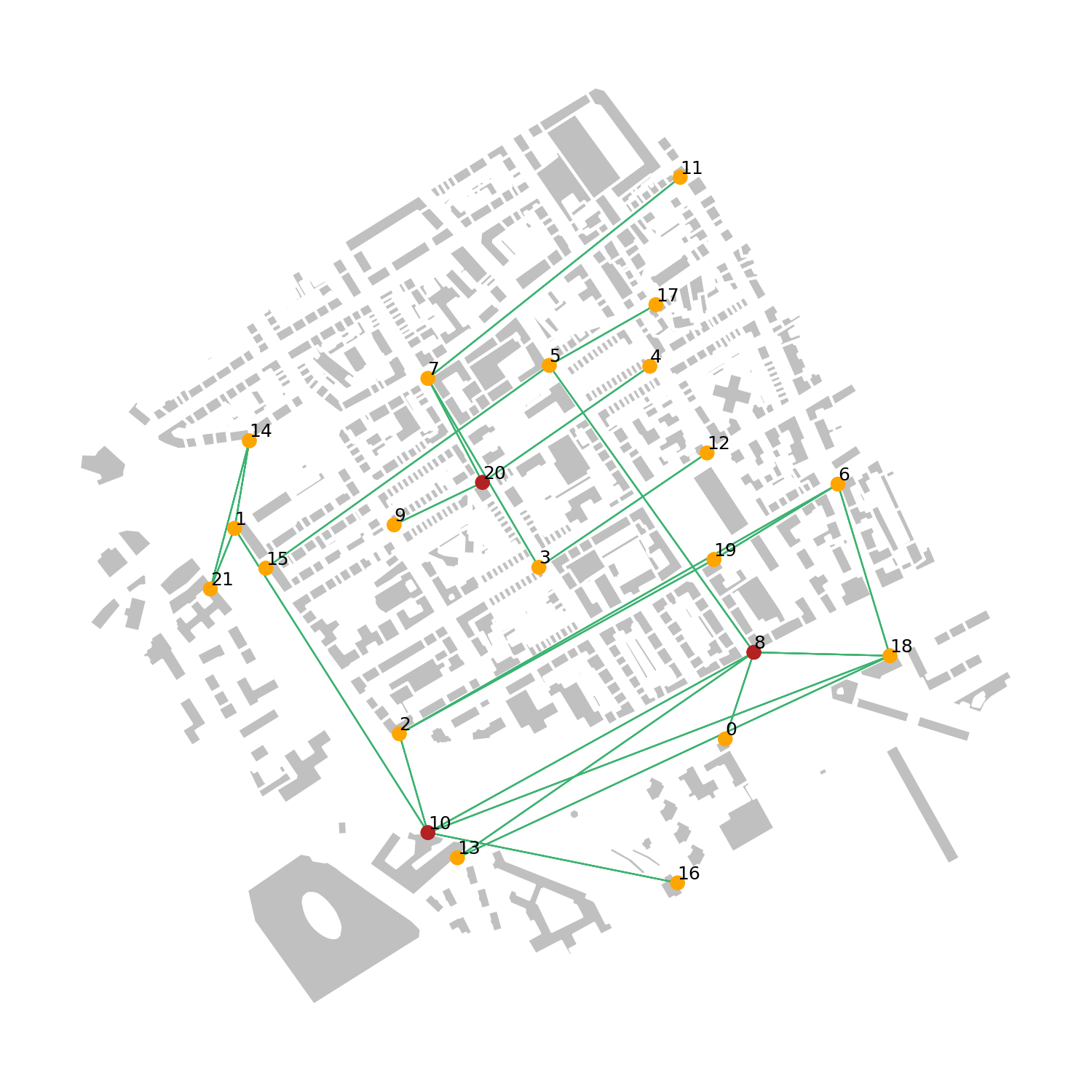}
    \caption{Realistic deployment scenario in Florence, Italy. Donors are represented in red, while IAB-Nodes are represented in yellow.}\label{fig:areas}
\end{figure}

\Gls{rtt} is measured when the network is unloaded, that is when there is no traffic flowing through the \gls{iab} network. As shown in Figure~\ref{fig:rtt_linear}, the first hop latency is around 11~ms. 
This value represents the base processing delay plus a small fixed propagation delay that is, however, the same for each hop. 
As the number of hops increases, the \gls{rtt} experiences a linear increase comparable with the first hop latency, as expected. 
This shows how the system does not introduce any spurious latency when the network is unloaded. 
Finally, the relatively higher \gls{rtt} standard deviation of the last hop (as represented by the whiskers in Figure~\ref{fig:rtt_linear}) suggests that multiple packet retransmissions are required. 

\subsection{Validation over realistic RF scenarios}

After having validated the system performance in a controlled environment, we move to more realistic urban scenarios, representing the typical deployment environment of an IAB network. 
We start by selecting a densely populated area of a city, from which we extract a precise 3D model of the environment. 
On top of this, we apply a coverage planning heuristic to find the optimal locations for the \gls{iab}-Nodes~\cite{gemmi2022cost}.
We then take advantage of a viewshed algorithm implemented on GPU and the above-mentioned 3D models to evaluate the \gls{los} between each pair of locations and to produce the so-called visibility graph \cite{gemmi2022properties}. 
Then, we characterize the propagation according to the 3GPP Urban Micro parametric model \cite{3gpp.38.901}, and we produce a tapered-delay representation of the communication channels, which Colosseum can then emulate. 
This process has been carried out for several European cities and four different scenarios are made available.\footnote{\url{https://colosseumneu.freshdesk.com/support/solutions/articles/61000303373-integrated-access-and-backhaul-scenarios}}

Motivated by the fact that \gls{iab} is unanimously considered as a key enabler of \gls{mmwave} \gls{ran}~\cite{cudak2021}, we are interested in providing an experimental solution that enables testing in such conditions. 
While Colosseum is not directly capable of operating at frequencies higher than 6~GHz, we can approximate these radio scenarios by reproducing the most relevant propagation characteristics of \glspl{mmwave}, namely the extremely directive transmissions through beamforming and the increased pathloss~\cite{kutty2015beamforming}. 
In particular, the pathloss between nodes that are not directly connected in the provided topologies has been set to infinite. The resulting suppression of inter-node interference might appear too ideal at first. 
However, this is compatible with the highly directive transmissions typical of \gls{mmwave}, where interference in static conditions (i.e., as in a backhaul IAB topology) can be practically neglected~\cite{fiore2022}. A more refined \gls{mmwave} channel emulation will be subject of future extensions.
In addition, since Colosseum's channel emulation happens in base-band, we can apply arbitrary pathloss independently of the radio frequency employed during the experiments. 
Thanks to this flexibility, we could compute pathloss for a carrier frequency of 28~GHz and apply them to \gls{los} links. 
Nonetheless, the scenarios made available to the Colosseum community are available for both 3.6~GHz and 28~GHz, both with and without inter-node interference suppression.

For the experimental evaluation presented in this work, we have selected a scenario based on the city of Florence, Italy. 
Figure~\ref{fig:areas} shows both the urban layout and the IAB deployment, which is extended over $0.7~\text{km}^2$ and comprises 21 nodes (3 of which are IAB-Donors). 
To determine which nodes are going to become \gls{iab}-Donors, we have applied the group closeness centrality metric \cite{bavelas1950communication} to the visibility graph. This centrality metric selects $k$ nodes such that their distance to all the other nodes is minimized. Then, we have determined the IAB topology as a Shortest-Path Forest computed over the visibility graph of the area with the well-known Dijktra's Algorithm. 
Similar to what has been done for the previous analysis, we characterize the throughput and latency at each hop in the network. In this case, however, the different link lengths cause performance variations in the per-hop throughput and latency. 
As such, we employ box plots to synthetically describe the network performance statistics in Figure~\ref{fig:florence_meas}. 
In particular, consider Figure~\ref{fig:florence_meas_dl}. 
Here the bottom and top edges of each box represent the first and third quartile of the downlink throughput measurements taken at all the different hops in the scenario. 
Similarly, the central marks indicate the median, and the whiskers represent the extreme data points. 
The plotted values indicate how the realistic pathloss introduced in the study scenario causes lower performance than the ideal case previously analyzed, independently of the considered hop. 
The same can be noted for the uplink throughput, as shown in Figure~\ref{fig:florence_meas_ul}. 
In both cases, the decreasing per-hop throughput trend is conserved. 
However, the throughput variability is the same for the two transmission directions. 
This is because, as opposed to the ideal scenario, the link length now represents the main performance-determining factor. 
This is testified by the significant distance between the first and third quartile of the first hop in both downlink and uplink throughput, which is consistent with the high variance of the first hop length in the topology of study. 
As for the second and third hop, the relatively closer quartiles are motivated by lower link length variations for these hops in the considered topology. 
Finally, the upper whiskers represent the performance of the shortest links, giving a further characterization of the system performance in this realistic scenario. 

Figure~\ref{fig:florence_meas_rtt} shows the \gls{rtt} statistic through the same plotting technique. 
Differently from the throughput, the latency is not affected by the link length variations in the considered scenario for the first two hops. 
Additionally, the \gls{rtt} increase at hops 1 and 2 is consistent with the one experienced in the controlled scenario. 
On the other hand, the high \gls{rtt} variance of the third and last hop suggests a high probability of requiring retransmissions along the IAB path. 

\section{Conclusions}
\label{sec:conclusions}
In this work, we have discussed the motivations and challenges of integrating \gls{iab} with \gls{oran}. 
On this matter, we have proposed possible architecture extensions that enable dynamic control and data collection over \gls{5g} \gls{iab} networks through \gls{oran} intelligent controllers. 
We have implemented the proposed integrated architecture and packaged it into the first publicly available experimental framework enabling at-scale testing and prototyping of \gls{oran}-based solutions applied to \gls{iab} networks. 
The system comprises all the software components required to establish end-to-end connectivity, plus custom-developed E2 and O1 agents that allow software-defined IAB-Nodes to be \gls{oran}-compliant. 
The framework is designed to be open and accessible and can be deployed over \gls{cots} hardware. 
We numerically validated the framework exploiting the large-scale testing capabilities of Colosseum, showing the system's performance over both an ideal linear topology and more sophisticated realistic deployments. 
Finally, the framework has been packaged and released into OpenRAN Gym and is available to the research community. 

\footnotesize
\bibliographystyle{IEEEtran}
\bibliography{wowmom23}
\end{document}